\documentclass{iopjournal}

\newcommand{\gtrsim}{\mathrel{\raise0.3ex\hbox{$>$}
\mkern-14mu\lower0.6ex\hbox{$\sim$}}}

\newcommand{\lesssim}{\mathrel{\raise0.3ex\hbox{$<$}
\mkern-14mu\lower0.6ex\hbox{$\sim$}}}

\begin{document}

\articletype{Paper} 

\title{Intrinsic Spectral Curvature from Finite-Cycle Transport at Relativistic Shocks}

\author{Ji-Hoon Ha$^{1, *}$\orcid{0000-0001-7670-4897}}

\affil{$^1$Korea Astronomy and Space Science Institute, Daejeon, Republic of Korea}

\affil{$^*$Author to whom any correspondence should be addressed.}

\email{jhha@kasi.re.kr}

\keywords{Relativistic shocks, Particle acceleration, Transport processes}

\begin{abstract}
Power-law spectra are a central prediction of shock acceleration and are commonly associated with asymptotic scale invariance under diffusive transport. In finite relativistic shocks, strong anisotropy and limited residence times may restrict the number of effective shock crossings before the many-cycle diffusive limit is established. This work develops a reduced finite-cycle framework in which particle energization is described by discrete shock-crossing mappings, while downstream transport is encoded through an energy-dependent return probability. In this formulation, the local spectrum is controlled by the competition between the mean energy gain per cycle and the probability of surviving to the next cycle. A systematic decrease of the return probability with energy then produces intrinsic spectral curvature as a consequence of transport-limited cycle survival. The energy dependence of the return probability is estimated from the competition between magnetic deflection, downstream advection, and finite shock lifetime, yielding a characteristic steepening scale determined by macroscopic source parameters. For fiducial parameters relevant to compact blazar emission regions, the steepening scale lies below the ultimate acceleration cutoff, so that curvature can appear before the terminal maximum energy is reached. These results point to a pre-asymptotic finite-cycle limit of relativistic shock transport in which non-power-law spectra can arise from the limited survival of repeated shock-crossing cycles.
\end{abstract}

\section{Introduction}
Relativistic shocks are widely invoked as sites of high-energy particle acceleration in a variety of astrophysical environments, including blazar jets, gamma-ray bursts, and radio galaxies \cite{Summerlin2012,Baring2016}.
In many theoretical treatments, particle acceleration at shocks is described within the framework of diffusive shock acceleration (DSA) \cite{Bell1978,Blandford1978,Drury1983,Blandford1987}, which assumes frequent pitch-angle scattering and near-isotropy of the particle distribution in the local fluid frame.
Under these conditions, the transport of energetic particles can be approximated by a diffusion equation, leading to quasi-universal power-law spectra.

At relativistic shocks, however, the conditions required to realize an asymptotic many-cycle diffusive regime may not always be satisfied, particularly in compact and transient systems.
Strong anisotropy in particle pitch-angle distributions, relativistic shock speeds, and limited residence times near the shock front can substantially reduce the number of effective shock crossings a particle undergoes before downstream removal.
Recent particle-in-cell (PIC) simulations have demonstrated that relativistic shocks are capable of producing nonthermal particle populations, although the efficiency and extent of acceleration depend sensitively on magnetic obliquity, upstream magnetization, and
turbulence \cite{Sironi2009,Sironi2013}.
In particular, superluminal or highly magnetized configurations can strongly suppress upstream return and inhibit the development of an extended many-cycle diffusive power-law tail \cite{Sironi2015,Tomita2022,Bresci2023,vanMarle2024,Mahlmann2025}.
Even in favorable configurations, PIC simulations often find that only a limited fraction of particles participates in sustained acceleration \cite{Mahlmann2025, Sironi2010,Ligorini2021,Kirk2023}.

Shock acceleration efficiency can also depend sensitively on the surrounding turbulence and transport conditions.
For example, pre-existing turbulence may substantially modify the efficiency of diffusive shock acceleration and the resulting particle transport properties \cite{Ha2025}, while kinetic simulations of relativistic shocks show that turbulence can either suppress or revive repeated shock crossings depending on the magnetic geometry and plasma
state \cite{Bresci2023}.
These results suggest that particle acceleration in realistic plasma environments is not governed solely by an idealized universal diffusive regime, but can remain strongly modulated by finite and structured transport conditions.

Internal shocks in flat-spectrum radio quasars (FSRQs) provide a prime example of such environments \cite{Spada2001, Mimica2007}.
They arise from collisions between relativistic plasma shells with different bulk Lorentz factors, persist only for a finite dynamical time, and are embedded in intense external radiation fields associated with the broad-line region and dusty torus \cite{Mimica2007, Pian2007}.
In these systems, particle acceleration is expected to occur under conditions where downstream advection, finite shock duration, and strong anisotropy play a central role \cite{Sironi2010, Ligorini2021, Kirk2023}.
As a result, the number of effective shock-crossing cycles is limited, and the accelerated particle spectrum need not be universal.
Compact black-hole coronae may provide another example of such finite acceleration environments.
Recent studies have suggested that nonthermal particle populations can be produced in hot magnetized coronae through shocks, turbulence, and related transport processes in spatially confined accretion flows \cite{Romero2010,Ly2026}.
Because these systems are dynamically evolving and characterized by strong magnetic fields, rapid transport, and finite residence times, particle acceleration may likewise remain strongly influenced by finite-cycle transport effects.

Recent numerical studies of relativistic jets have further emphasized that particle acceleration often takes place in finite, dynamically evolving flows characterized by shocks, turbulence, and velocity shear \cite{Seo2021,Seo2023,Cerutti2023,Perucho2023}.
Such nonlinear structures arise naturally through the dissipation of kinetic energy in relativistic outflows and may play an important role in the production of ultrahigh-energy cosmic rays \cite{Seo2023,seo2025}.
In realistic jet environments, particle transport is therefore governed by spatially structured and time-dependent conditions rather than by
idealized steady-state configurations.
However, how particle acceleration operates when the number of effective acceleration cycles remains finite under such structured conditions remains theoretically unclear.

Despite significant progress in understanding particle acceleration at relativistic shocks from first-principles simulations and transport studies \cite{Achterberg2001}, an important question remains: \emph{how does particle acceleration proceed when asymptotic diffusive transport is not efficiently realized in finite relativistic systems?} 
In particular, if repeated shock crossings are limited by strong anisotropy, rapid downstream advection, or the finite lifetime of the acceleration region, the establishment of asymptotic diffusive transport may become inefficient. 
While previous studies have demonstrated that relativistic shocks can sustain highly anisotropic particle distributions, most theoretical treatments still focus on regimes in which repeated scattering ultimately produces a quasi-steady power-law spectrum through many-cycle transport. 
A complementary reduced framework is therefore useful for describing particle energization in finite relativistic systems, where acceleration may be influenced by a limited number of effective shock-crossing cycles.

This work develops a reduced finite-cycle framework for particle acceleration at relativistic shocks.  Previous studies have demonstrated that strong anisotropy can limit the applicability of the diffusion approximation and strongly influence shock-crossing transport \cite{Achterberg2001}.
More recent work has further emphasized that the acceleration rate, limiting energy, and spectral form can depend sensitively on the transport regime, magnetic-field geometry, and finite confinement of the system \cite{Kirk2023, Perri2022,Huang2023}.
The aim of the present approach is not to introduce a new energization mechanism, but to formulate a reduced description of relativistic shock acceleration in terms of finite-cycle survival.
Rather than taking the asymptotic many-cycle power-law state as the reference solution, the present model focuses on cases in which only a limited number of effective shock-crossing cycles can be completed.
Particle energization is described as a sequence of discrete shock-crossing mappings, while downstream transport is encoded through an energy-dependent return probability that captures, in a cycle-averaged sense, the combined effects of magnetic deflection, downstream advection, and finite shock lifetime.
In this picture, the accelerated spectrum is not modeled as a phenomenological modification of an already-established diffusive power law.  
Rather, the scale-free transport state may not be fully established when repeated cycle survival is progressively suppressed with energy.
Within this finite-cycle framework, the model yields closed-form expressions for the particle spectrum, acceleration timescale, and characteristic steepening energy, and illustrates the resulting transport-induced spectral curvature for fiducial internal shocks relevant to compact blazar emission regions.

\begin{figure*}[t]
\includegraphics[width=\linewidth]{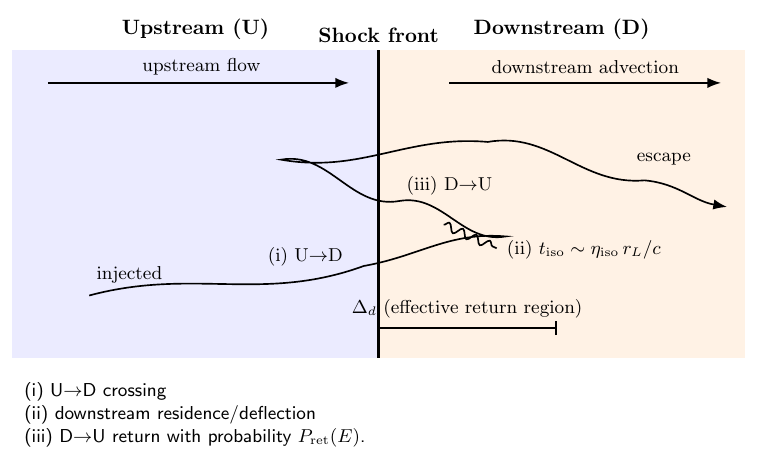}
\caption{Schematic illustration of finite-cycle acceleration at a relativistic shock. This figure is intended to show the physical sequence of shock crossing, downstream residence, and possible return, and is not a quantitative model calculation.
Particles undergo discrete cycles: (i) upstream-to-downstream crossing, (ii) downstream residence and directional deflection by magnetic irregularities on a timescale $t_{\rm iso}\sim \eta_{\rm iso} r_L/c$, and (iii) downstream-to-upstream return with probability $P_{\rm ret}(E)$.
Finite downstream extent and finite shock lifetime limit the number of effective cycles; particles that fail to return are advected downstream and escape.}
\label{fig:f1}
\end{figure*}

\section{Finite-cycle shock acceleration}

The physical mechanism considered in this work is illustrated schematically in Figure~\ref{fig:f1}.
Particles injected upstream undergo a sequence of discrete shock-crossing cycles consisting of downstream transmission, finite downstream residence, and a possible return to the shock.
While the underlying energization process remains similar to that commonly invoked in DSA, the transport regime considered here differs from the standard asymptotic diffusive picture.
Rather than assuming arbitrarily many effective scattering cycles and near-isotropy at all energies, particle transport is described in terms of finite directional deflection within a magnetized downstream flow.
In this regime, particles undergo magnetic deflection through gyro-motion and pitch-angle scattering in the downstream magnetic field over a characteristic timescale $t_{\rm iso}\sim \eta_{\rm iso} r_L/ c$, which represents an effective directional-return timescale rather than complete stochastic isotropization. The parameter $\eta_{\rm iso}$ is an effective angular-scattering parameter, analogous to the mean-free-path factor used in diffusive transport models, with $\lambda_{\rm scatt}\sim \eta_{\rm iso}r_L$. In terms of an effective pitch-angle diffusion coefficient, $t_{\rm iso}$ corresponds to the order-of-magnitude time required for pitch-angle deflection sufficient to allow downstream-to-upstream return. Thus, smaller $\eta_{\rm iso}$ corresponds to stronger turbulence and more efficient angular scattering, whereas larger $\eta_{\rm iso}$ corresponds to weaker turbulence, more ordered magnetic fields, or stronger anisotropic transport.

During this interval, particles are simultaneously subject to downstream advection and the finite lifetime of the shock, both of
which act as removal channels.
Only particles redirected toward the shock before being removed from the acceleration region can complete additional cycles, naturally
producing an energy-dependent return probability.
As the particle energy increases, the growing Larmor radius lengthens the directional-deflection time and progressively suppresses repeated
shock crossings.
As a result, the transport process no longer approaches the many-cycle scale-free limit underlying standard DSA.
The number of effective return cycles therefore remains transport-limited rather than asymptotically large, giving rise to intrinsic spectral curvature through finite-cycle transport.

\subsection{Finite-cycle shock-crossing kinematics}
The model considers a planar relativistic shock. In the shock rest frame, the upstream and downstream fluids move along the shock normal with velocities $\beta_{\rm u}$ and $\beta_{\rm d}$, respectively. 
The corresponding relative velocity and Lorentz factor are
\begin{equation}
\beta_{\rm rel}=\frac{\beta_{\rm u}-\beta_{\rm d}}{1-\beta_{\rm u}\beta_{\rm d}},
\qquad
\Gamma_{\rm rel}=(1-\beta_{\rm rel}^2)^{-1/2}.
\end{equation}
Particles are characterized by momentum $p$ and pitch-angle cosine $\mu=\cos\theta$ measured in the local fluid frame.  
For a particle transmitted from the upstream to the downstream region, the momentum and pitch-angle cosine transform as
\begin{eqnarray}
p_{\rm d}^{(n)} &=& \Gamma_{\rm rel}\,p_{\rm u}^{(n)}
\left(1-\beta_{\rm rel}\mu_{\rm u}^{(n)}\right), \\
\mu_{\rm d}^{(n)} &=& 
\frac{\mu_{\rm u}^{(n)}-\beta_{\rm rel}}
{1-\beta_{\rm rel}\mu_{\rm u}^{(n)}}.
\end{eqnarray}
Here, the superscript $(n)$ denotes quantities associated with the $n$th shock-crossing cycle.
For a particle returning from downstream to upstream,
\begin{eqnarray}
p_{\rm u}^{(n+1)} &=& 
\Gamma_{\rm rel}\,p_{\rm d}^{(n)}
\left(1+\beta_{\rm rel}\mu_{\rm d}^{(n)}\right), \\
\mu_{\rm u}^{(n+1)} &=& 
\frac{\mu_{\rm d}^{(n)}+\beta_{\rm rel}}
{1+\beta_{\rm rel}\mu_{\rm d}^{(n)}}.
\end{eqnarray}
The admissible pitch-angle domains for repeated shock crossings become progressively narrower as $\Gamma_{\rm rel}$ increases, leading to increasingly anisotropic transport and reduced cycle survival probability.

\subsection{Transport without asymptotic diffusion: cycle operator}

Rather than adopting a continuous diffusion description, acceleration is described as a discrete-cycle evolution in phase space. 
Let $f_{\rm u}^{(n)}(p,\mu)$ and $f_{\rm d}^{(n)}(p,\mu)$ denote the upstream and downstream particle distributions associated with the $n$th shock-crossing cycle, respectively. 
A single cycle consists of: (i) upstream-to-downstream transmission, (ii) downstream residence with finite directional deflection, and (iii) downstream-to-upstream return or escape.

Introducing a downstream return kernel $\mathcal{R}_{{\rm d}}(\mu_{\rm ret}|\mu_{\rm d},p_{\rm d})$ and escape probability $P_{{\rm esc},{\rm d}}(p_{\rm d},\mu_{\rm d})$, the upstream distribution evolves as
\begin{equation}
f_{\rm u}^{(n+1)}(p,\mu)
=
\int d p_0\,d\mu_0\;
\mathcal{K}(p,\mu|p_0,\mu_0)\,
f_{\rm u}^{(n)}(p_0,\mu_0),
\label{eq:cycle_operator}
\end{equation}
where $\mathcal{K}$ represents the effective cycle operator obtained by composing the Lorentz-transformed shock-crossing maps with downstream return and escape processes. 
The corresponding downstream distribution is generated through the upstream-to-downstream crossing map and subsequent downstream transport during the $n$th cycle.
Probability conservation implies
\begin{equation}
\int d\mu_{\rm ret}\,
\mathcal{R}_{{\rm d}}(\mu_{\rm ret}|\mu_{\rm d},p_{\rm d})
+
P_{{\rm esc},{\rm d}}(p_{\rm d},\mu_{\rm d})
=
1.
\end{equation}
The escaping population accumulated over repeated cycles is formally given by
\begin{equation}
N_{\rm esc}(p)
=
\sum_{n\ge0}
\int d\mu\;
P_{{\rm esc},{\rm d}}(p,\mu)\,
f_{\rm d}^{(n)}(p,\mu).
\end{equation}

In the scale-free many-cycle limit with energy-independent return probability, stationary power-law eigenfunctions may emerge. 
In finite relativistic shocks where the return probability depends on energy, the asymptotic scale invariance underlying standard diffusive acceleration can be weakened, and the resulting spectrum can deviate from a universal power law.

\subsection{Leading-order cycle model}

The full cycle operator in equation~(\ref{eq:cycle_operator}) contains the detailed angular redistribution, return statistics, and momentum mapping associated with relativistic shock-crossing transport.
To obtain closed-form results, it is replaced by a leading-order cycle-averaged closure.
The purpose of this reduction is not to reproduce the full pitch-angle structure of relativistic shock acceleration, but to isolate the spectral effect of finite-cycle anisotropic transport, namely the progressive suppression of repeated cycle survival.
After angular averaging, the two quantities that directly control the cycle-by-cycle construction of the accelerated population are the return probability $P_{\rm ret}$ and the mean momentum amplification factor $g$.
The former fixes the surviving normalization from one cycle to the next, while the latter fixes the mean displacement in momentum per completed cycle.
Angular and momentum-space correlations neglected in this closure are therefore not assumed to be unimportant in general, but are treated as higher-order corrections to the leading survival-and-amplification description.
Such correlations may broaden the spectrum, smooth the transition region, or modify the detailed curvature profile, but they would not be expected to remove the finite-cycle steepening as long as the effective return probability decreases systematically with energy.
In this sense, the model should be interpreted as a reduced analytic closure of the general transport operator rather than as a complete kinetic description.

Repeated applications of the cycle operator formally generate
\begin{equation}
f_{\rm u}^{(n)}
=
\mathcal{K}^n f_{\rm u}^{(0)} .
\end{equation}
In the asymptotic scale-free limit, the transport evolution may be dominated by stationary eigenmodes of $\mathcal{K}$.
Rather than solving the full angular eigenvalue problem, the following analysis uses this cycle-averaged closure to obtain a minimal analytic description of finite-cycle spectral formation. In general, the exact operator may contain higher-order correlations between pitch angle, momentum gain, residence time, and return probability. Such terms would produce momentum-space broadening, phase-space memory, and possible correlations between the gain in a given cycle and the probability of completing the next return. These effects are not assumed to be absent in realistic relativistic shocks, but are not solved explicitly in the present analytic model. The closure is most appropriate when the leading spectral behavior is governed by the cycle-averaged survival probability and mean logarithmic energy gain, rather than by a broad gain distribution or strong gain-return correlations. Higher-order couplings are expected to modify the detailed curvature profile or smooth the transition region, while the leading finite-cycle steepening remains controlled by the energy dependence of the effective return probability. A full treatment of these higher-order phase-space couplings would require an angle-dependent transport calculation, Monte Carlo modeling, or kinetic simulations.

It is important to specify how relativistic-shock anisotropy enters the cycle-averaged formulation used here. In the full cycle operator, anisotropy is retained explicitly through the pitch-angle variable $\mu$, the Lorentz-transformed shock-crossing maps, and the downstream return kernel ${\cal R}_{\rm d}(\mu_{\rm ret}|\mu_{\rm d},p_{\rm d})$. Only particles whose downstream pitch angles are redirected into the range allowing downstream-to-upstream return can complete another acceleration cycle. Thus, before angular averaging, anisotropy directly controls the fraction of particles that can return to the shock. In the leading-order closure adopted below, this angular dependence is not solved explicitly, but is coarse-grained into the effective return probability $P_{\rm ret}$ and the directional-return time $t_{\rm iso}$. The downstream return kernel is therefore approximated by an angularly averaged redistribution with total weight $P_{\rm ret}$,
\begin{equation}
{\cal R}_{\rm d}(\mu_{\rm ret}|\mu_{\rm d},p_{\rm d})
=
\frac{P_{\rm ret}}{2}\Theta(1-|\mu_{\rm ret}|),
\qquad
P_{{\rm esc},{\rm d}}=1-P_{\rm ret}.
\label{eq:minimal_kernel}
\end{equation}
This reduced prescription intentionally integrates over the detailed pitch-angle structure of the downstream distribution in order to retain only the cycle-averaged survival statistics relevant for finite-cycle transport. Strong anisotropy, ordered magnetic fields, or oblique and superluminal shock geometries reduce the accessible return fraction and increase the effective directional-return time by making pitch-angle deflection less efficient. Conversely, strong turbulence broadens the pitch-angle distribution, enhances the probability of return, and lowers the effective angular-scattering parameter $\eta_{\rm iso}$. In the same reduced sense, upstream turbulence, magnetic obliquity, magnetization, and subluminal or superluminal field geometry are not modeled explicitly, but are absorbed into the angular distribution entering the shock-crossing maps, the effective return probability, and the angular-scattering parameter $\eta_{\rm iso}$. Strong upstream turbulence would generally favor repeated crossings, whereas ordered, strongly magnetized, or superluminal configurations would suppress return and increase the effective directional-return time. Therefore, anisotropy is not ignored in the reduced formulation, but enters through the cycle-survival probability and the timescale controlling directional return.

After this angular averaging, the cycle operator reduces schematically to the
mapping
\begin{equation}
p_{\rm u}^{(n+1)} \approx g\,p_{\rm u}^{(n)},
\qquad
N_{n+1}\approx P_{\rm ret}\,N_n ,
\label{eq:reduced_cycle_map}
\end{equation}
where
\begin{equation}
g\equiv
\left\langle
\frac{p_{\rm u}^{(n+1)}}{p_{\rm u}^{(n)}}
\right\rangle
\approx \Gamma_{\rm rel}^2 .
\label{eq:gdef}
\end{equation}
The estimate $g\approx\Gamma_{\rm rel}^2$ is used only as a leading-order characterization of the cycle-averaged energy gain. In realistic relativistic shock acceleration, the gain per cycle is stochastic and depends on the pitch-angle distribution, shock-crossing angles, magnetic geometry, and scattering history. Thus, the quantity $g$ in the present reduced model should be interpreted as an effective mean, more precisely logarithmically averaged, amplification factor rather than as a deterministic gain applied identically to all particles. For mildly relativistic internal shocks, this estimate should therefore be understood as an order-of-magnitude characterization of the average cycle gain rather than a universal asymptotic result. Fluctuations in the cycle-to-cycle gain would broaden the distribution in energy and smooth the resulting spectral curvature, but they would not remove the finite-cycle steepening as long as the return probability decreases systematically with energy. A broad gain distribution, or strong correlations between the gain and the return probability, could modify the detailed high-energy tail and would require a full angular transport or Monte Carlo treatment. In the present reduced closure, the intrinsic curvature derived below is controlled primarily by the energy dependence of $P_{\rm ret}$ rather than by the precise numerical value of the mean gain factor.

If $P_{\rm ret}$ is independent of momentum, the reduced cycle map gives,
after $n$ completed cycles,
\begin{equation}
N_n=N_0P_{\rm ret}^n,
\qquad
p_n=p_0g^n .
\end{equation}
Eliminating $n$ then yields the familiar scale-free result
\begin{equation}
N(>p)\propto p^{\ln P_{\rm ret}/\ln g},
\end{equation}
and the corresponding differential spectrum is
\begin{equation}
\frac{dN}{dp}\propto p^s,
\qquad
s=\frac{\ln P_{\rm ret}}{\ln g}-1 .
\label{eq:slope_const}
\end{equation}
Thus, in the energy-independent limit, the reduced operator admits a scale-free power-law solution whose slope is determined by the cycle survival probability relative to the mean momentum gain. This provides an explicit consistency check with the asymptotic DSA-like limit. When both the cycle survival probability and the mean gain are independent of momentum, and when many shock-crossing cycles can be completed, the cycle evolution becomes statistically self-similar. No preferred momentum scale is then introduced, and the reduced cycle model recovers the usual many-cycle power-law behavior. The intrinsic curvature discussed below appears only when this self-similarity is broken by an energy-dependent return probability.

When the return probability depends on momentum, however, the surviving
population is no longer described by a simple geometric progression.
Instead,
\begin{equation}
N_n
=
N_0
\prod_{i=0}^{n-1}
P_{\rm ret}(p_i),
\qquad
p_i=p_0g^i .
\label{eq:cycle_product}
\end{equation}
Taking the logarithm gives
\begin{equation}
\ln N_n
=
\ln N_0
+
\sum_{i=0}^{n-1}
\ln P_{\rm ret}(p_0g^i).
\label{eq:log_product}
\end{equation}
For a slowly varying return probability, this discrete expression defines a local cycle-averaged spectral slope.  
Since $d\ln p/dn=\ln g$, the local cumulative slope may be written as
\begin{equation}
\frac{d\ln N(>p)}{d\ln p}
\approx
\frac{\ln P_{\rm ret}(p)}{\ln g}.
\label{eq:local_cumulative_slope}
\end{equation}
The corresponding local differential spectral index is therefore
\begin{equation}
s(p)
\approx
\frac{\ln P_{\rm ret}(p)}{\ln g}-1 .
\label{eq:slope}
\end{equation}
Equation~(\ref{eq:slope}) should therefore be interpreted as a local
cycle-averaged spectral index rather than as an exact global power-law
solution.  If $P_{\rm ret}$ is independent of momentum, it reduces to the
scale-free result in equation~(\ref{eq:slope_const}).  If $P_{\rm ret}$ decreases systematically with energy, the effective cycle-survival eigenvalue varies with energy, the scale-free many-cycle limit is broken, and the spectrum develops intrinsic curvature.

Equation~(\ref{eq:slope}) may also be interpreted from an operator perspective. 
The cycle mapping defined by equation~(\ref{eq:cycle_operator}) acts as a linear transport operator on the phase-space distribution. 
In the scale-free limit with energy-independent $P_{\rm ret}$, repeated applications of this operator yield stationary power-law solutions governed by its dominant eigenvalue. 
When $P_{\rm ret}$ becomes energy dependent, however, the effective eigenvalue varies with energy, breaking the asymptotic scale invariance and naturally producing intrinsic spectral curvature.

\subsection{Relation to energy-dependent escape descriptions}

It is useful to clarify the relation between the present finite-cycle formulation and more conventional energy-dependent escape, finite-age, or leaky-box descriptions \cite{Drury1983,Blandford1987,Kirk2023,Perri2022,Huang2023,Malkov2001}. Energy-dependent escape is a standard ingredient in many shock-acceleration models. The purpose of the present formulation is not to introduce escape as a new effect, but to specify a different level of description. In conventional escape-limited or leaky-box treatments, particle acceleration is usually described by a continuous transport equation for an already coarse-grained particle population. Escape then enters as a loss term, often through a characteristic escape time $t_{\rm esc}(E)$, while finite source lifetime or confinement determines the terminal energy. Such descriptions are appropriate when many acceleration steps occur and a continuous description in energy space is justified.

By contrast, the present approach starts from the discrete shock-crossing process itself. The return probability $P_{\rm ret}(E)$ is the conditional probability that a particle which has entered the downstream region is redirected toward the shock and completes the next downstream-to-upstream return. It is therefore a cycle-completion probability rather than a phenomenological global escape time. The surviving population after $n$ completed cycles is written as
\begin{equation}
    N_n = N_0 \prod_{i=0}^{n-1} P_{\rm ret}(E_i),
    \qquad
    E_{i+1}=gE_i,
\end{equation}
so that the spectrum is determined by the product of cycle-survival probabilities before taking any continuum limit.

The connection to leaky-box descriptions can nevertheless be made in the many-cycle limit. If $\Delta t_{\rm cyc}$ denotes the mean duration of one acceleration cycle, the discrete survival probability may be mapped onto an effective escape rate as
\begin{equation}
t_{\rm esc}^{-1}(E) \approx -{\ln P_{\rm ret}(E)\over \Delta t_{\rm cyc}},
\end{equation}
while the mean acceleration rate may be written as
\begin{equation}
t_{\rm acc}^{-1} \approx {\ln g\over \Delta t_{\rm cyc}} .
\end{equation}
Under this continuum approximation, the finite-cycle formulation reduces to an energy-dependent escape model. The regime emphasized here, however, is the pre-asymptotic regime in which the number of effective shock-crossing cycles remains limited and the scale-free diffusive eigenstate is not fully established. In this case, spectral curvature arises because the cycle-survival eigenvalue itself varies with energy. The curvature is therefore not introduced as a modification of an already established DSA power law, but appears during the construction of the accelerated population through finite-cycle survival.

This distinction also separates the characteristic steepening scale from the terminal finite-age or confinement cutoff. The energy at which $t_{\rm iso}(E)\sim t_{\rm rem}$ marks the onset of inefficient cycle survival, whereas the maximum energy is determined separately by the available acceleration time or confinement condition. Thus, the finite-cycle steepening can occur below the ultimate cutoff energy.

\section{Energy-dependent return probability and spectral curvature}

\subsection{Physical estimate of $P_{\rm ret}(E)$}
A downstream particle completes an additional shock-crossing cycle only if
it remains within the effective downstream return region long enough to be
redirected toward the shock.  The return process can therefore be interpreted
as a finite-time survival problem: a particle must survive downstream removal
for at least the characteristic directional-deflection time.

The model includes two independent removal channels: downstream advection and finite shock lifetime. Downstream advection removes particles from the effective
return region on a timescale
\begin{equation}
t_{\rm adv}\sim \frac{\Delta_{{\rm d}}}{u_{{\rm d}}},
\label{eq:tadv}
\end{equation}
where $\Delta_{{\rm d}}$ is the effective downstream return region and
$u_{{\rm d}}$ is the downstream flow speed in the shock frame.  Throughout
this section, primed quantities denote values measured in the local comoving
downstream frame.  The finite lifetime of the acceleration region introduces
a second removal time,
\begin{equation}
t_{\rm life}\sim t_{\rm dyn}',
\label{eq:tlife}
\end{equation}
where $t_{\rm dyn}'$ denotes the characteristic dynamical lifetime of the
acceleration region.

In a reduced transport description, these two processes are treated as
statistically independent removal channels.  If each removal process is
approximated as memoryless on the timescale relevant for a single cycle, the
survival probability $S(t)$ satisfies
\begin{equation}
\frac{dS}{dt}
=
-\left(
\frac{1}{t_{\rm adv}}+\frac{1}{t_{\rm life}}
\right)S .
\label{eq:survival_equation}
\end{equation}
With $S(0)=1$, this gives
\begin{equation}
S(t)
=
\exp\!\left[
-\,t
\left(
\frac{1}{t_{\rm adv}}+\frac{1}{t_{\rm life}}
\right)
\right]
\equiv
\exp\!\left(-\frac{t}{t_{\rm rem}}\right),
\label{eq:survival_probability}
\end{equation}
where the effective removal time is
\begin{equation}
t_{\rm rem}^{-1}
=
t_{\rm adv}^{-1}+t_{\rm life}^{-1}.
\label{eq:trem_def}
\end{equation}

The remaining ingredient is the time required for a downstream particle to
be redirected toward the shock.  In the finite-cycle transport regime, this
directional-deflection time is parameterized as
\begin{equation}
t_{\rm iso}\sim \eta_{\rm iso}\,\frac{r_L}{c},
\qquad
r_L(E)=\frac{E}{qB'},
\label{eq:tiso}
\end{equation}
where $r_L$ is the Larmor radius, $q$ is the particle charge, and $B'$ is the
magnetic-field strength measured in the local comoving downstream frame.
The parameter $\eta_{\rm iso}$ phenomenologically accounts for finite angular
deflection efficiency.  In this work, $t_{\rm iso}$ should be understood as
an effective directional-return time rather than as a full isotropization
time in the diffusive sense.

A particle can return to the shock only if it survives removal for at least the directional-return time $t_{\rm iso}$. More generally, if $\psi_{\rm rem}(t,E)$ denotes the probability density for removal from the effective downstream return region, the return probability can be written as the survival probability
\begin{equation}
P_{\rm ret}(E) \sim \int_{t_{\rm iso}(E)}^\infty \psi_{\rm rem}(t,E)dt .
\end{equation}
This expression states that a particle contributes to the next acceleration cycle only when its downstream residence time exceeds the time required for directional deflection back toward the shock. The exponential form used below corresponds to the special case in which downstream advection and finite shock lifetime act as memoryless removal processes during a single cycle. In that case,
\begin{equation}
\psi_{\rm rem}(t) = t_{\rm rem}^{-1}\exp(-t/t_{\rm rem}),
\end{equation}
and therefore
\begin{equation}
P_{\rm ret}(E) \sim S[t_{\rm iso}(E)]
= \exp[-t_{\rm iso}(E)/t_{\rm rem}] .
\end{equation}
Using $t_{\rm rem}^{-1}=t_{\rm adv}^{-1}+t_{\rm life}^{-1}$ and $t_{\rm iso}\approx \eta_{\rm iso} r_L/c$, this gives
\begin{equation}
P_{\rm ret}(E) \sim
\exp\left[
-\eta_{\rm iso}{E\over qB'c}
\left({1\over t_{\rm adv}}+{1\over t_{\rm life}}\right)
\right].
\label{eq:Pret_energy}
\end{equation}
Equation~(\ref{eq:Pret_energy}) should therefore be regarded as the leading-order survival probability associated with finite-cycle transport under memoryless downstream removal, rather than as an imposed spectral ansatz. It decreases monotonically with energy because particles with larger Larmor radii require longer times to be redirected toward the shock.

The subsequent spectral curvature does not rely on the exponential form itself. Alternative removal-time distributions would lead to different functional forms of $P_{\rm ret}(E)$ and hence to different detailed curvature profiles. For example, a broader removal-time distribution would produce a more gradual steepening, whereas a narrower or more sharply truncated removal-time distribution would lead to a more abrupt turnover. However, the central requirement for intrinsic curvature is simply that the cycle-survival probability decreases systematically with energy. In that case, the effective cycle-survival eigenvalue becomes energy dependent, the scale-free many-cycle limit is broken, and a non-power-law spectrum is produced. Thus, the exponential expression adopted here should be interpreted as the simplest analytically transparent closure for finite-cycle survival, not as a unique assumption required for transport-induced spectral curvature.

The memoryless approximation should therefore be understood as a local hazard-rate closure over one acceleration cycle, not as a statement that realistic downstream residence times must be exactly exponential. In a more detailed shock model, $\psi_{\rm rem}(t,E)$ may depend on the downstream flow structure, turbulence, magnetic geometry, and finite size of the simulation or source region. Such effects would modify the detailed functional form of $P_{\rm ret}(E)$, but they would not remove the finite-cycle steepening as long as the probability of completing another return decreases with energy.

\begin{figure*}[t]
\includegraphics[width=\linewidth]{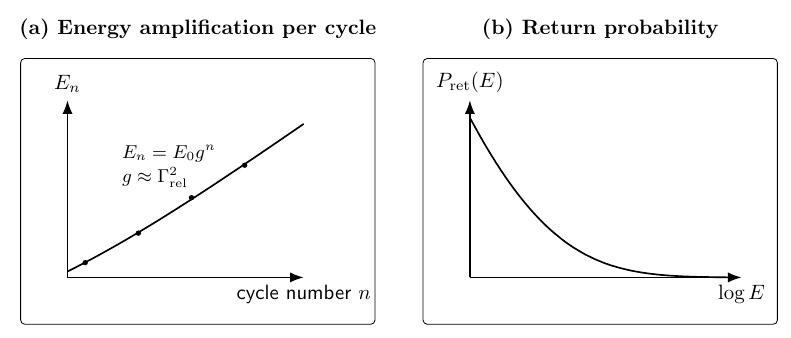}
\caption{Schematic illustration of the origin of finite-cycle spectral steepening. Panel (a) shows the idealized multiplicative energy gain per cycle, while panel (b) illustrates the qualitative decrease of the return probability with energy. The curves are schematic and are intended to visualize the mechanism rather than to represent a direct numerical solution.}
\label{fig:f2}
\end{figure*}

The physical origin of this steepening is illustrated schematically in Figure~\ref{fig:f2}.  Panel (a) shows that particles gain energy multiplicatively through successive shock-crossing cycles, $E_n = E_0 g^n$, with a nearly constant amplification factor determined primarily by the relativistic shock kinematics.  Panel (b) illustrates that the return probability decreases systematically with energy because the directional-deflection time increases with the Larmor radius. The combination of approximately constant mean energy gain and progressively suppressed cycle survival weakens the scale-free character of the many-cycle diffusive limit and can produce intrinsic spectral curvature. In this reduced model, the resulting steepening is not imposed by radiative losses or by an externally prescribed high-energy cutoff, but arises from the finite-cycle transport and escape prescription.

\begin{figure}[t]
\includegraphics[width=\linewidth]{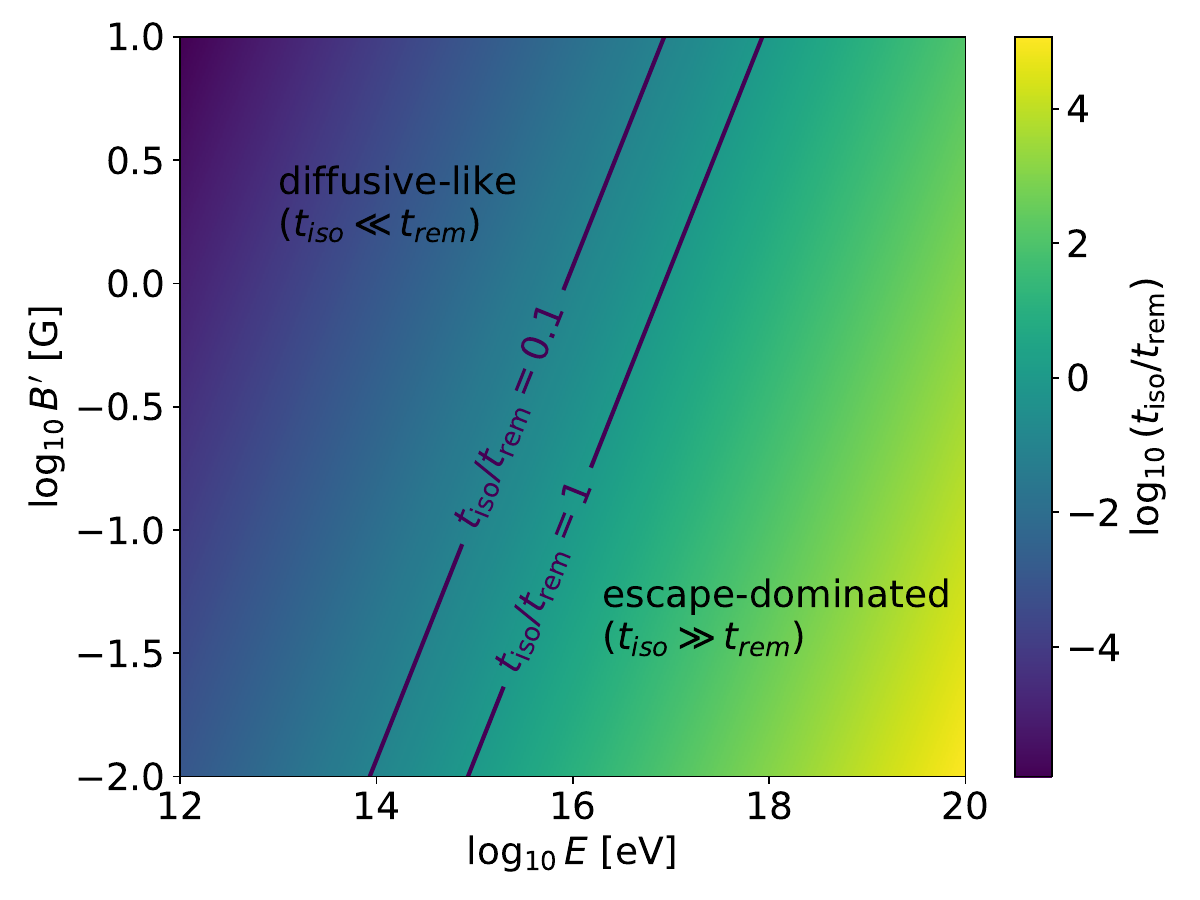}
\caption{
Model-derived regime map in the $(E,B')$ plane showing the ratio
$t_{\rm iso}/t_{\rm rem}$ for fiducial internal-shock parameters
$t_{\rm adv}=1.8\times10^{5}\,\mathrm{s}$,
$t_{\rm life}=2.0\times10^{5}\,\mathrm{s}$,
and $\eta_{\rm iso}=10$.
The effective removal time is
$t_{\rm rem}=(t_{\rm adv}^{-1}+t_{\rm life}^{-1})^{-1}$.
Contours denote $t_{\rm iso}/t_{\rm rem}=0.1$ and $1$,
separating diffusion-like transport from finite-cycle escape.}
\label{fig:f3}
\end{figure}

\begin{figure*}[t]
\includegraphics[width=\linewidth]{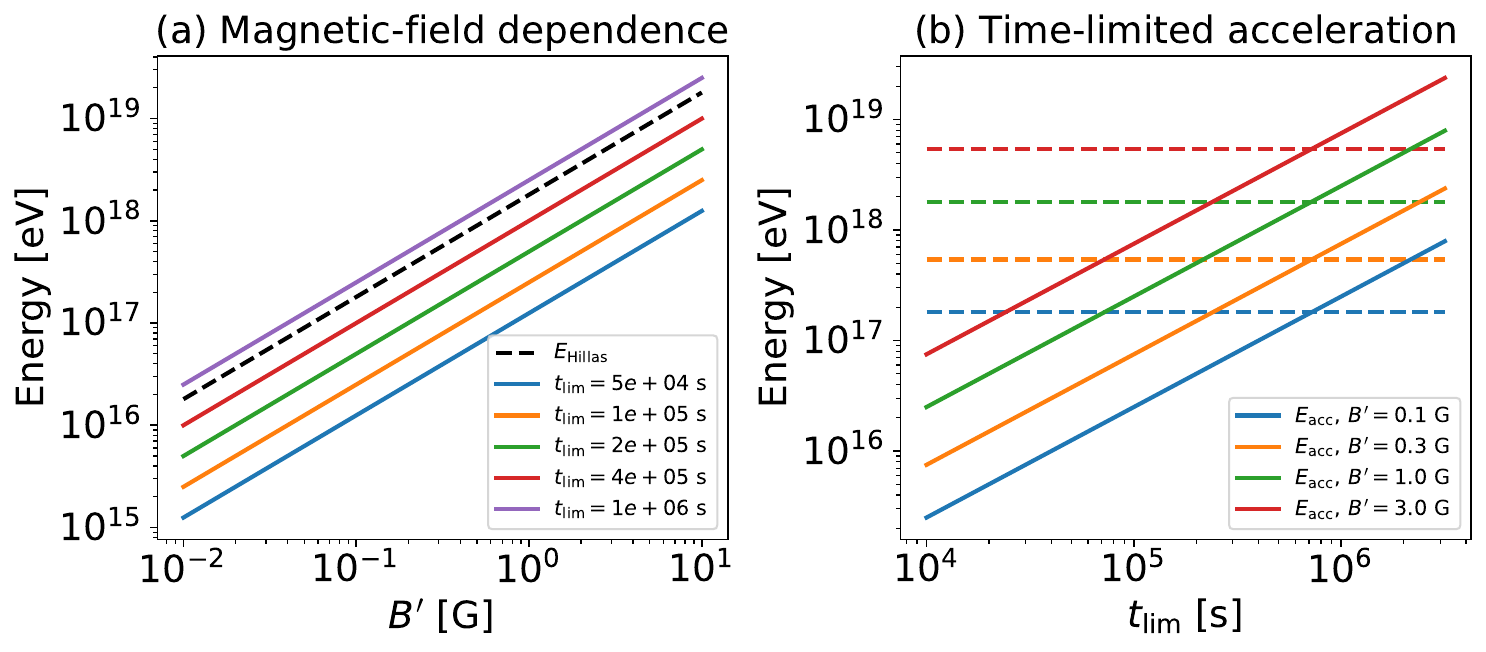}
\caption{
Model-derived comparison of the confinement (Hillas) limit and
the time-limited acceleration energy for a fiducial
internal-shock configuration
($R'=6\times10^{15}\,$cm, $\eta_{\rm acc}=10$).
(a) Magnetic-field dependence.
The dashed black line shows the Hillas limit,
$E_{\rm Hillas}\propto B'$.
Colored curves denote the time-limited energy
$E_{\rm acc}\propto B' t_{\rm lim}$
for several values of $t_{\rm lim}$.
(b) Explicit $t_{\rm lim}$ dependence for different
magnetic-field strengths.
Solid lines show $E_{\rm acc}$,
while dashed horizontal lines indicate the corresponding
$E_{\rm Hillas}$ values for the same $B'$.
For sufficiently short $t_{\rm lim}$,
$E_{\rm acc}<E_{\rm Hillas}$,
demonstrating that finite acceleration time
can provide a more restrictive upper bound
than confinement.
}
\label{fig:f4}
\end{figure*}

\begin{figure*}[t]
\includegraphics[width=\linewidth]{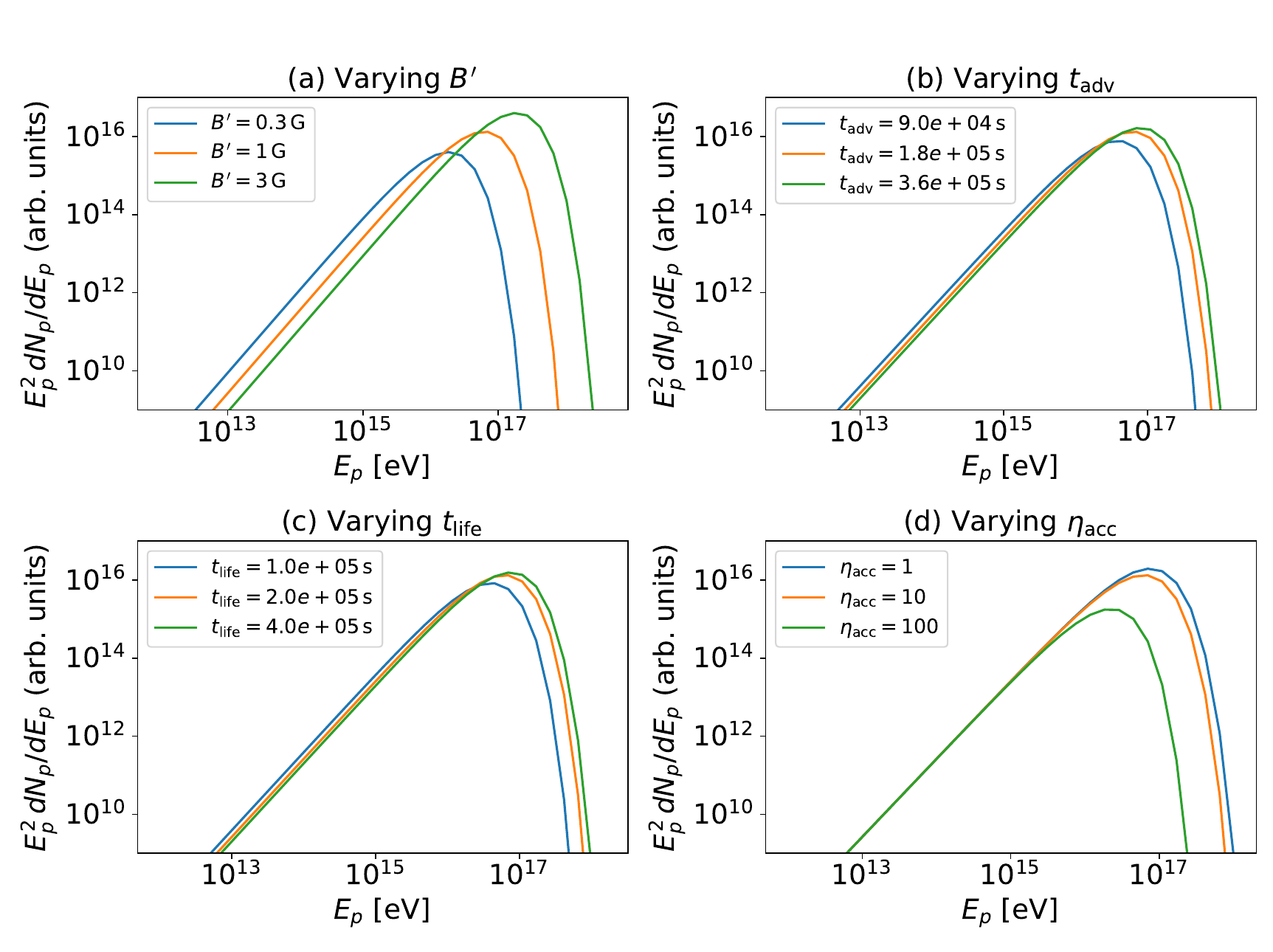}
\caption{Proton spectra from the reduced finite-cycle model for a fiducial internal-shock setup.
Reference parameters: $\Gamma_f=30$, $\Gamma_s=15$ ($\Gamma_{\rm rel}\approx1.25$), $t_{\rm var}=10^4\,{\rm s}$, $\delta=20$ ($R'\approx6\times10^{15}\,{\rm cm}$), $B'=1\,{\rm G}$, $t_{\rm adv}\approx1.8\times10^{5}\,{\rm s}$, $t_{\rm life}\approx2.0\times10^{5}\,{\rm s}$, $\eta_{\rm iso}=10$, and $\eta_{\rm acc}=10$.
Panels show variations with (a) $B'$, (b) $t_{\rm adv}$, (c) $t_{\rm life}$, and (d) $\eta_{\rm acc}$.
Curvature develops gradually via energy-dependent escape, whereas the terminal cutoff reflects confinement and time-limited acceleration.}
\label{fig:f5}
\end{figure*}

The ratio $t_{\rm iso}/t_{\rm rem}$ provides a natural quantitative diagnostic of the transport regime. When $t_{\rm iso}/t_{\rm rem} \ll 1$, particles are typically redirected toward the shock before being removed and may undergo many effective crossings, supporting a diffusion-like many-cycle description. Conversely, once $t_{\rm iso}/t_{\rm rem}\gtrsim 1$, particles are likely to escape before completing additional cycles, suppressing repeated returns and leading to a finite-cycle escape-dominated regime. In this sense, the characteristic steepening scale corresponds physically to the transition $t_{\rm iso}/t_{\rm rem}\sim 1$.

Figure~\ref{fig:f3} shows the resulting regime map in the $(E,B')$ plane for fiducial internal-shock parameters. The boundary defined by $t_{\rm iso}\sim t_{\rm rem}$ marks an energy-dependent transition between diffusion-like transport and escape-dominated acceleration. For magnetic fields typical of compact blazar emission regions, the transition occurs in the PeV--EeV range, illustrating that finite-cycle behavior can emerge naturally from the competition between particle deflection and macroscopic removal processes without invoking extreme microphysical conditions. This criterion also provides a simple way to interpret numerical simulations: weakly magnetized or strongly turbulent shocks that sustain repeated returns and extended nonthermal tails correspond to $t_{\rm iso}/t_{\rm rem}\ll 1$ or transitional values, whereas highly magnetized, superluminal, or strongly anisotropic shocks with inefficient upstream return are expected to lie closer to the $t_{\rm iso}/t_{\rm rem}\gtrsim 1$ finite-cycle regime.

\subsection{Energy-dependent spectral index and steepening scale}

Substituting equation~(\ref{eq:Pret_energy}) into
equation~(\ref{eq:slope}) yields an energy-dependent spectral index,
\begin{equation}
s(E)=\frac{\ln P_{\rm ret}(E)}{\ln g}-1
=
-1-\frac{\eta_{\rm iso}}{\ln g}\frac{E}{qB'c}
\left(\frac{1}{t_{\rm adv}}+\frac{1}{t_{\rm life}}\right).
\label{eq:sE_final}
\end{equation}
It is convenient to rewrite this expression as
\begin{equation}
s(E)=-1-\frac{E}{E_*},
\qquad
E_*\equiv
\frac{qB'c\,\ln g}{\eta_{\rm iso}\left(t_{\rm adv}^{-1}+t_{\rm life}^{-1}\right)}.
\label{eq:Estar_def}
\end{equation}

The characteristic scale $E_*$ admits a direct physical interpretation.
It corresponds to the energy at which the directional deflection time
becomes comparable to the effective removal time,
$t_{\rm iso}(E_*) \sim t_{\rm rem}$,
where $t_{\rm rem}^{-1}=t_{\rm adv}^{-1}+t_{\rm life}^{-1}$.
Below $E_*$, particles are typically redirected toward the shock before
being removed, allowing multiple shock crossings and maintaining a
spectrum close to a power law.
Above $E_*$, the survival probability decreases exponentially with
energy, progressively suppressing repeated returns and producing a
systematic steepening of the spectrum.
The scale $E_*$ therefore marks a transport-induced transition rather
than a radiative or confinement-imposed cutoff.

Because the spectral index varies linearly with $E/E_*$, the resulting
spectrum does not exhibit a sharp break.
Instead, the local slope increases continuously with energy,
implying genuine intrinsic curvature.
This behavior is a direct consequence of the exponential form of
$P_{\rm ret}(E)$ and reflects the finite number of effective
shock-crossing cycles available in compact relativistic systems.

In realistic astrophysical environments, radiative cooling can also
modify particle spectra.
However, the curvature isolated in the present model arises from the acceleration and transport physics of finite-cycle escape, without explicitly including energy-loss processes.
The present analysis therefore isolates the intrinsic spectral
curvature produced by transport-limited acceleration.
This suggests that deviations from scale-free power laws can arise from the acceleration and transport process itself, even before radiative losses are included.

\subsection{Fiducial internal-shock estimate}
As an illustrative application, consider internal shocks relevant to compact blazar emission zones.
Representative shell Lorentz factors $\Gamma_f=30$ and $\Gamma_s=15$ are adopted, corresponding to a relative Lorentz factor $\Gamma_{\rm rel}\approx1.25$ \cite{Spada2001, Mimica2007}. 
Within the present leading-order cycle model, this gives a characteristic energy amplification factor $g\approx \Gamma_{\rm rel}^2\approx1.56$ ($\ln g\approx0.446$).
The fiducial parameters adopted here correspond to mildly relativistic internal shocks typical of compact blazar emission regions.
In the present framework, the significance of finite-cycle transport does not depend on extremely large cycle gains, but on the inability of the system to realize an asymptotic many-cycle transport regime.
With variability time $t_{\rm var}=10^4~{\rm s}$ and Doppler factor $\delta\approx20$, the comoving size is $R'\sim \delta c t_{\rm var}\approx 6\times10^{15}~{\rm cm}$.
Let $\Delta_{{\rm d}}=\xi R'$ with $\xi=0.3$, $u_{{\rm d}}\approx c/3$, and $t_{\rm life}\sim R'/c$.
Then
\begin{equation}
t_{\rm adv}\approx 1.8\times10^5~{\rm s},
\qquad
t_{\rm life}\approx 2.0\times10^5~{\rm s}.
\end{equation}
For $B'=1~{\rm G}$ and $\eta_{\rm iso}=10$, equation~(\ref{eq:Estar_def}) gives
\begin{equation}
E_*\approx 8.5\times10^{16}~{\rm eV}\approx 85~{\rm PeV}.
\end{equation}
Thus the spectrum steepens mildly at PeV energies and more strongly above tens of PeV.
For illustration,
\begin{eqnarray}
s(1~{\rm PeV}) &\sim& -1.03,\\ \nonumber
s(10~{\rm PeV}) &\sim& -1.26,\\ \nonumber
s(100~{\rm PeV}) &\sim& -3.64.
\end{eqnarray}
The steepening scale $E_*$ depends primarily on $B'$ and the macroscopic removal rates $t_{\rm rem}^{-1}$.
Keeping the same kinematic setup but varying $B'$ within a typical FSRQ range, one finds
\begin{eqnarray}
E_* &\sim& 26~{\rm PeV}\quad (B'=0.3~{\rm G}), \\ \nonumber
E_* &\sim& 260~{\rm PeV}\quad (B'=3~{\rm G}),
\end{eqnarray}
showing that spectral curvature in the PeV--EeV decade is a natural outcome of finite-cycle acceleration for compact, transient internal-shock parameters.

The parameter dependence follows directly from equation~(\ref{eq:Estar_def}). The steepening scale increases with $B'$ and with the effective removal time $t_{\rm rem}$, but decreases with increasing $\eta_{\rm iso}$. Thus, stronger magnetic fields or longer downstream residence times allow particles to complete additional cycles up to higher energies, delaying the onset of curvature. By contrast, less efficient angular scattering, represented by larger $\eta_{\rm iso}$, shifts the finite-cycle steepening to lower energies. The removal time is controlled by both downstream advection and finite shock lifetime through $t_{\rm rem}^{-1}=t_{\rm adv}^{-1}+t_{\rm life}^{-1}$; therefore, the shorter of $t_{\rm adv}$ and $t_{\rm life}$ usually provides the dominant limitation. If downstream advection is rapid, $t_{\rm adv}$ controls the curvature scale, whereas if the shock lifetime is shorter, $t_{\rm life}$ becomes the limiting timescale.

The fiducial value $\eta_{\rm iso}=10$ adopted here represents moderately efficient, but non-Bohm, angular scattering. For resonant scattering, a rough order-of-magnitude estimate gives $\eta_{\rm iso}\sim (B/\delta B)^2$, up to geometric factors and details of the turbulence spectrum. Thus, $\eta_{\rm iso}\sim 10$ corresponds to a moderate turbulence amplitude, $\delta B/B\sim 0.3$, rather than to either the Bohm limit or an extremely weak-scattering regime. This choice is a physically plausible fiducial case for a relativistic downstream region in which magnetic turbulence is present, but a compressed or partially ordered field component still limits rapid angular isotropization. Stronger turbulence would move the system toward $\eta_{\rm iso}\sim 1$ and shift the curvature scale to higher energies, whereas more ordered, weakly turbulent, or superluminal configurations could correspond to $\eta_{\rm iso}\sim 10$--$100$ or larger and would move the onset of finite-cycle steepening to lower energies.

The present calculation should also be understood as a quasi-steady or time-averaged closure. In a genuinely transient shock, the magnetic field strength, angular-scattering efficiency, downstream advection time, and effective shock lifetime may evolve with time. The return probability and the characteristic steepening energy would then become time dependent, and the observed or escaping spectrum would represent a superposition of finite-cycle spectra produced at different evolutionary stages of the shock. Slow evolution compared with the characteristic cycle time can be approximated by the quasi-static prescription adopted here, whereas rapid evolution may broaden the curvature, shift the apparent steepening energy, or introduce additional time-dependent spectral features. A fully time-dependent treatment with evolving injection, shock structure, turbulence, and adiabatic losses is beyond the scope of the present reduced model, but $t_{\rm life}$ provides a first-order representation of the finite duration of the acceleration episode.

The absolute maximum energy is constrained by particle confinement,
commonly expressed through the Hillas limit \cite{Hillas1984},
\begin{equation}
E_{\rm Hillas}\sim qB'R'
\approx 1.8\times10^{18}\left(\frac{B'}{1~{\rm G}}\right)\left(\frac{R'}{6\times10^{15}\ {\rm cm}}\right)\ {\rm eV},
\end{equation}
and by time-limited acceleration.
Parameterizing the acceleration time as $t_{\rm acc}\sim \eta_{\rm acc} r_L/c$ gives
\begin{equation}
E_{\rm acc}\sim \frac{qB'c\,t_{\rm lim}}{\eta_{\rm acc}},
\qquad
t_{\rm lim}=\min(t_{\rm adv},t_{\rm life}),
\end{equation}
or numerically
\begin{equation}
E_{\rm acc}\approx
1.8\times10^{17}
\left(\frac{B'}{1~{\rm G}}\right)
\left(\frac{t_{\rm lim}}{2\times10^{5}\ {\rm s}}\right)
\left(\frac{\eta_{\rm acc}}{10}\right)^{-1}
{\rm eV}.
\end{equation}
Hence $E_{\max}=\min(E_{\rm Hillas},E_{\rm acc})$ typically lies in the $10^{17}$--$10^{18}\,$eV range for these parameters, while curvature already appears at $E\sim E_*$.

Figure~\ref{fig:f4} compares the confinement limit
$E_{\rm Hillas}$ and the time-limited acceleration energy
$E_{\rm acc}$ for representative internal-shock parameters.
Panel (a) shows the magnetic-field dependence.
Both limits scale linearly with $B'$, but
$E_{\rm acc}$ also depends on the available acceleration time.
For shorter $t_{\rm lim}$, the acceleration constraint
lies significantly below the Hillas bound,
indicating that finite acceleration time, rather than spatial confinement,
can control the attainable terminal energy.
Panel (b) illustrates the explicit $t_{\rm lim}$ dependence.
While $E_{\rm Hillas}$ is independent of $t_{\rm lim}$ and therefore
appears as horizontal dashed lines,
$E_{\rm acc}\propto B't_{\rm lim}$ increases linearly with time.
For sufficiently small $t_{\rm lim}$,
$E_{\rm acc}<E_{\rm Hillas}$, placing the system in an
acceleration-limited regime.
Only for sufficiently large $t_{\rm lim}$ does the confinement limit become dominant.

Figure~\ref{fig:f5} illustrates how finite-cycle escape and global
maximum-energy constraints jointly shape the proton spectrum.
In all panels, the spectrum first develops a gradual curvature,
set by the escape-driven steepening scale $E_*$,
well before reaching the ultimate cutoff.
Varying $B'$ primarily shifts both $E_*$ and the maximum energy,
consistent with their linear dependence on the magnetic field strength.
Changes in $t_{\rm adv}$ and $t_{\rm life}$ modify the residence time
in the downstream region and therefore regulate the efficiency of
repeated shock crossings, leading to measurable but comparatively
moderate shifts of the high-energy turnover.
By contrast, panel (d) demonstrates that the acceleration parameter
$\eta_{\rm acc}$ directly controls the acceleration timescale and can
significantly lower the attainable maximum energy when the acceleration
process becomes inefficient.
For typical FSRQ internal-shock parameters, the hierarchy
$E_* \ll E_{\max}$ implies that the spectrum can become curved before
the terminal cutoff is reached.
hus, Figure~\ref{fig:f5} highlights the physical separation between
transport-induced curvature, controlled by finite-cycle return
probability, and the ultimate cutoff, controlled by confinement and
time-limited acceleration.

The spectra shown in Figure~\ref{fig:f5} are computed for protons and should be regarded as a rigidity-normalized baseline for the finite-cycle transport effect. The same formalism can be extended qualitatively to heavier ions because the directional-return time is controlled primarily by the particle rigidity. Since $t_{\rm iso}\propto r_L/c$ and $r_L\propto E/(ZB')$ for a relativistic ion of charge $Ze$, the characteristic finite-cycle steepening scale is expected to shift approximately as
\begin{equation}
E_{*,Z}\approx Z E_{*,p},
\end{equation}
provided that $\eta_{\rm iso}$ and the macroscopic removal times are comparable for different species. A mixed composition would therefore produce a superposition of species-dependent curvature scales and could broaden the transition in the all-particle spectrum. This simple rigidity scaling, however, does not include composition-dependent injection, hadronic interactions, photodisintegration, or species-dependent scattering. In photon-rich relativistic sources, such as blazar jets or gamma-ray burst internal shocks, nuclear interactions may further modify or suppress the high-energy heavy-ion component. A full treatment of mixed composition therefore requires a dedicated time-dependent composition calculation and is beyond the scope of the present reduced model. The proton spectra presented here should be interpreted as an illustrative baseline for transport-induced curvature rather than as a complete composition model.

\section{Summary and discussion}
This work developed a finite-cycle framework for acceleration at relativistic shocks, in which energization proceeds through discrete shock-crossing mappings and downstream transport is encoded by a return probability.
In a minimal cycle model, the spectrum is controlled by $P_{\rm ret}$ and the mean gain per cycle $g$.
When $P_{\rm ret}(E)$ decreases with energy, the resulting spectrum exhibits intrinsic curvature rather than a universal power law.
Estimating $P_{\rm ret}(E)$ from the competition between magnetic deflection, advection, and finite shock lifetime yields a characteristic steepening scale $E_*$ that depends on macroscopic source properties.
For fiducial internal-shock parameters relevant to compact blazar emission zones, curvature emerges at tens of PeV while the ultimate cutoff is controlled independently by confinement and time-limited acceleration, implying $E_* \ll E_{\max}$ in typical regimes.
This suggests that non-power-law spectra may arise from acceleration and transport effects, even before radiative cooling or propagation effects are included.
More broadly, this result suggests that the scale invariance often assumed in asymptotic shock-acceleration theory may be modified in compact or transient systems where only a limited number of effective shock-crossing cycles can be completed.

For the fiducial parameters considered here, the confinement and time-limited constraints yield maximum energies in the $10^{17}$--$10^{18}$\,eV range.
While a detailed treatment of composition and escape is beyond the scope of this work, these energies approach the regime commonly associated with ultrahigh-energy cosmic rays.
In this context, the intrinsic spectral curvature derived above may influence the injection spectrum of hadrons from compact relativistic shocks, independently of radiative or propagation effects.

The finite-cycle interpretation is not restricted to blazar internal shocks. Similar conditions may occur in other compact or relativistic environments where the acceleration region has a finite lifetime and downstream-to-upstream return is not guaranteed. In gamma-ray burst internal shocks, for example, the comoving dynamical time is short, the particle distribution is highly anisotropic, and repeated shock crossings may be limited by rapid expansion and downstream advection \cite{Rees1994, Pian2005}. In such systems, finite-cycle transport could introduce curvature into the accelerated hadron or lepton spectrum before the terminal maximum energy is reached. Pulsar-wind termination shocks provide another possible environment, although the geometry is more complex because of magnetic obliquity, striped-wind structure, reconnection, and strong downstream turbulence \cite{Kennel1984, Sironi2011}. In these systems, the present framework should be regarded as a reduced description of the competition between angular return and downstream removal rather than as a complete model of the shock structure.

It is also important to distinguish the transport-induced curvature discussed here from curvature produced by radiative losses. Synchrotron and inverse-Compton cooling introduce spectral breaks through energy losses after or during acceleration, with a characteristic scale determined by the comparison between the cooling time and the dynamical or escape time \cite{Kardashev1962,Blumenthal1970, Rybicki1986}. By contrast, the curvature in the present model arises from the acceleration process itself: the probability of completing additional shock-crossing cycles decreases as the directional-return time increases with particle energy. Thus, the relevant scale is controlled by angular transport, downstream advection, and finite shock lifetime rather than by the radiative cooling time. In realistic sources, both effects may operate simultaneously, and the observed spectrum would reflect the combined action of finite-cycle transport, radiative cooling, and source evolution. Observationally, cooling-induced breaks are expected to correlate primarily with magnetic-field or radiation energy density, whereas finite-cycle curvature should be more closely tied to transport conditions such as magnetic obliquity, turbulence level, source size, downstream advection, and shock lifetime. In hadronic applications, transport-induced curvature in the parent proton spectrum may also appear in secondary gamma-ray or neutrino spectra without requiring a corresponding electron synchrotron cooling break at the same energy. A detailed source-by-source separation of these effects requires time-dependent radiation modeling and is beyond the scope of the present work, but these differences indicate possible observational diagnostics.

A comparison with existing numerical studies helps clarify the physical regime represented by the present reduced model. Monte Carlo calculations of ultra-relativistic shock acceleration have shown that, when repeated shock crossings remain efficient and the scattering prescription is approximately scale free, the accelerated particles approach an asymptotic power-law with a nearly universal index \cite{Achterberg2001}. This corresponds to the energy-independent limit of the present reduced cycle model. In contrast, Monte Carlo studies of oblique and superluminal relativistic shocks show that the resulting spectra and acceleration efficiency depend sensitively on magnetic-field obliquity, the scattering prescription, and the turbulence level \cite{Summerlin2012,Baring2017}. These trends are consistent with the present interpretation: configurations that reduce the probability of downstream-to-upstream return, or increase the effective angular-deflection time, move the system away from the many-cycle limit and toward finite-cycle steepening. The present model does not attempt to reproduce the full angular Monte Carlo spectra, but provides a reduced analytic description of the same physical tendency, namely the progressive suppression of repeated shock-crossing survival when angular return becomes inefficient.

The present work isolates the transport physics of finite-cycle acceleration and provides a compact mechanism for generating non-power-law spectra in relativistic finite systems without assuming asymptotically efficient diffusive transport.
An immediate extension is to couple the curved parent-hadron spectra predicted here to secondary emission calculations (e.g., photohadronic neutrinos and accompanying electromagnetic cascades) in radiation-rich environments such as blazar jets, which is left for future work.

\ack{The numerical calculations were performed using computing resources provided by Korea Astronomy and Space Science Institute.}

\funding{This research received no external funding.}

\roles{Ji-Hoon Ha: Conceptualization, Methodology, Formal analysis, Investigation, Writing – original draft, Writing – review \& editing.}

\data{No new data were generated or analyzed in this study.}


\bibliographystyle{iopart-num}
\bibliography{ref}

\end{document}